\documentclass[a4paper,11pt]{article}
\usepackage{jcappub} 
\usepackage{lineno}
\usepackage{aas_macros}
\usepackage{subcaption}
\usepackage{graphicx}

\title{Novel Challenges in Tracking Self-Interacting Dark Matter Subhalos}

\author[a, 1]{Demao Kong,\note{Corresponding author}}
\author[a]{Hai-Bo Yu,}
\author[b]{Ethan O. Nadler,}
\author[c,d]{Philip Mansfield}
\author[e]{and Andrew Benson}
\emailAdd{dkong012@ucr.edu}
\emailAdd{haiboyu@ucr.edu}
\emailAdd{enadler@ucsd.edu}
\emailAdd{phil1@stanford.edu}
\emailAdd{abenson@carnegiescience.edu}

\affiliation[a]{Department of Physics and Astronomy, University of California, Riverside, CA 92521, USA}
\affiliation[b]{Department of Astronomy and Astrophysics, University of California, San Diego, CA 92093, USA}

\affiliation[c]{Kavli Institute for Particle Astrophysics \& Cosmology, P. O. Box 2450, Stanford University, Stanford, CA 94305, USA}

\affiliation[d]{SLAC National Accelerator Laboratory, Menlo Park, CA 94025, USA}

\affiliation[e]{Carnegie Science, 813 Santa Barbara Street, Pasadena, CA 91101, USA}

\abstract{Cosmological $N$-body simulations are among the primary tools for studying structure formation in the Universe. Analyses of these simulations critically depend on accurately identifying and tracking dark matter subhalos over time. In recent years, several new algorithms have been developed to improve the accuracy and consistency of subhalo tracking in cold dark matter (CDM) simulations. These algorithms should be revisited in the context of new physics beyond gravity, which can modify the evolution and final properties of subhalo populations. In this work, we apply the particle-tracking-based subhalo finder Symfind to velocity-dependent self-interacting dark matter (SIDM) simulations with large cross section amplitudes to assess the performance of particle-tracking methods beyond the CDM paradigm. We find that the core-particle-tracking technique, which is key to the success of these algorithms in CDM, does not always yield accurate results in SIDM. In particular, the interplay between dark matter self-interactions and tidal stripping can cause the diffusion of core particles to larger radii, leading particle-tracking-based algorithms to prematurely lose track of SIDM subhalos. For massive core-expansion subhalos and core-collapse subhalos that experience close or repeated pericentric passages, a significant fraction of core particles can be lost, and particle-tracking-based finders such as Symfind offer no clear advantage over traditional methods that rely on identifying phase-space overdensities. On the other hand, for subhalos with large pericentric distances or fewer, more distant passages, Symfind tends to outperform. These differences depend sensitively on the cross section amplitude and turnover velocity of the underlying SIDM model. We therefore recommend a hybrid approach that leverages the strengths of both techniques to produce complete and robust catalogs of core-expansion and core-collapse SIDM subhalos.}

\begin{document}
\maketitle
\flushbottom

\section{Introduction}
\label{sec:intro}

In the standard Lambda Cold Dark Matter ($\Lambda$CDM) cosmological model, dark matter (DM) halos originate from primordial density fluctuations. Large structures form hierarchically through the merging of numerous small halos. During this merging process, halos that are accreted into a larger halo still survive as self-bounded ``subhalos'' of their hosts~\cite{1997MNRAS.286..865T,1998ApJ...499L...5M, 1998MNRAS.300..146G,1999ApJ...522...82K, 2008MNRAS.391.1685S}. Predictions for DM substructure are crucial to test our current understanding of structure formation on small scales and to interpret observations of satellite galaxies, stellar streams, and strong-lensing systems. Importantly, the distribution and evolution of substructure are sensitive to the particle nature of DM, since physics beyond gravity can change subhalo abundances and density profiles \cite{Spergel:1999mh,2012MNRAS.423.3740V, 2013MNRAS.431L..20Z,2014MNRAS.439..300L,2016PhRvD..93l3527C,Sameie:2019zfo, Zavala:2019sjk,2021MNRAS.505.5327T,2021PhRvL.126i1101N,Nadler:2020ulu,Correa:2022dey,Shah:2023qcw,2023ApJ...949...67Y, 2023ApJ...958L..39N,Fischer:2023lvl,Despali:2025koj,Nadler:2025wra,Nadler:2025jwh}. Subhalos are therefore critical for testing DM on sub-galactic scales and addressing the diverse distribution of DM in galaxies, which is currently a prominent challenge to the $\Lambda$CDM framework~\citep{Salucci:2000ps,KuziodeNaray:2009oon,Kaplinghat:2015aga,2015MNRAS.452.3650O, Kamada:2016euw,Valli:2017ktb,2018MNRAS.473.4392S, 2019PhRvX...9c1020R,2019MNRAS.490..231K, 2019MNRAS.487.5272J, 2019MNRAS.487.1380G, 2019ApJ...887...94R, 2020MNRAS.495...58S,Kong:2022oyk,PinaMancera:2021wpc,ManceraPina:2024ybj, 2022NatAs...6..897S,Roberts:2024uyw,Zhang:2024ggu,Zhang:2024fib,Kong:2025irr}.

Due to the complexity of structure formation and its non-linear nature, cosmological N-body simulations remain a primary tool for understanding such processes. Even with increasing computational power and resolution over the years, N-body simulations still face numerical challenges. A particularly important effect is that numerical disruption can destroy subhalos prematurely or nonphysically, leading to an underestimation of subhalo abundances~\cite{1987ApJ...313..505W, 1994ApJ...433..468C, 1995MNRAS.273..295V, 2002ApJ...570..656J, 2003ApJ...584..541H, 2010ApJ...709.1138D, 2016MNRAS.457.1208H, 2017MNRAS.468..885V, 2017MNRAS.472..657J, 2019MNRAS.488.3143B, 2024MNRAS.533.3811D}. Another challenge is to correctly identify and relate subhalos across simulation snapshots, especially for subhalos that have been heavily tidally stripped~\cite{2011MNRAS.415.2293K, 2013MNRAS.435.1618K, 2024ApJ...970..178M, 2024MNRAS.533.3811D}. Both questions are blended into each other for subhalos, and as a result, the drivers of subhalo disruption in N-body simulations are often ambiguous.

Several recent studies have developed different software tools and numerical techniques to address these problems. Of particular relevance for this paper are friends-of-friends algorithms that group particles belonging to distinct phase-space overdensities, and particle-tracking algorithms that follow some subset of a subhalo's particles forward in time, e.g., after infall into a larger host \cite{1985ApJ...292..371D, 1995ApJ...454....1S, 2006MNRAS.371..537W, 2010MNRAS.404.1111G, 2010ApJ...710..903M, 2012MNRAS.427.2437H,2013ApJ...762..109B, 2017ApJS..231....5D, 2018MNRAS.474..604H,2019PASA...36...21E, 2019ApJS..245...16H, 2021ApJ...913..109S, 2021MNRAS.506.2871S, 2021ApJS..252...19H, 2023OJAp....6E..24K, 2024ApJ...970..178M, 2024MNRAS.533.3811D, Moreno:2025nps}. Multiple particle-tracking-based halo finders have been developed based on a similar underlying algorithm that first associates particles with subhalos at a certain time, using a phase-space friends-of-friends algorithm, and then tracks particles forward to determine descendant subhalos in later snapshots. Compared to algorithms that solely rely on phase-space information, particle-tracking approaches recover more subhalos and are able to follow subhalos longer, although the exact performance varies between different simulations and halo finders~\cite{2012MNRAS.427.2437H, 2018MNRAS.474..604H, 2017ApJS..231....5D, 2024MNRAS.533.3811D, 2024ApJ...970..178M, 2021MNRAS.506.2871S}. However, even though the particle-tracking technique generally improves the accuracy of DM substructure predictions, its fidelity should be revisited in scenarios involving DM physics beyond gravity. In these scenarios, the tidal evolution of subhalos can differ from CDM expectations, and all particle-tracking-based algorithms involve certain numerical choices and assumptions that are not always physically motivated.

In this work, we therefore apply the particle-tracking-based finder Symfind~\cite{2024ApJ...970..178M} to track the evolution of subhalos in the self-interacting DM (SIDM) framework and scrutinize its performance compared to phase-space-based halo finders in both SIDM and CDM, using simulation data from the \textsc{SIDM Concerto} compilation~\cite{Nadler:2025jwh}. SIDM has been proposed as an alternative to CDM that explains the diverse DM distribution in galaxies (see \cite{Tulin:2017ara} for a review). Self-interactions between DM particles can modify the inner regions of DM (sub)halos relative to CDM through a process known as gravothermal evolution. The thermalization process creates isothermal cores that are low-density, called the core-expansion stage~\cite{Spergel:1999mh,Burkert:2000di,Dave:2000ar,
Yoshida:2000uw,2000ApJ...543..514K, 2002ApJ...581..777C, 2012MNRAS.423.3740V, 2013MNRAS.430..105P, 2013MNRAS.430...81R, 2013MNRAS.431L..20Z, 2015MNRAS.453...29E,Robertson:2016qef,Fischer:2022rko}. As the evolution continues, the SIDM core will eventually transport heat outward, leading to a rising DM density, called the core-collapse stage~\cite{2002ApJ...568..475B,Koda:2011yb,Essig:2018pzq,Nishikawa:2019lsc,Feng:2020kxv,2022JCAP...09..077Y,Gilman:2022ida,Outmezguine:2022bhq,Yang:2022zkd,2025PhRvD.111f3001Z,Jiang:2025jtr,Feng:2025rzf,Roberts:2025poo,Gurian:2025zpc,Kamionkowski:2025fat}. We will show that self-interactions between DM particles can kick tightly-bound particles in the inner regions of a halo out to larger radii or out of the subhalo entirely. In the presence of a tidal field, the initial inner particles (often referred to as ``core particles'') can be stripped away and dissociated from the subhalo, leading to an incorrect determination of subhalo disruption when using particle-tracking algorithms that rely upon these core particles. 

Compared to the phase-space-based subhalo finder Rockstar~\cite{2013ApJ...762..109B}, we find that Symfind can spuriously lose track of SIDM subhalos due to the unbinding of core particles. This primarily happens for subhalos with high infall masses that are in the core-expansion stage of gravothermal evolution and are subsequently heavily stripped. On the other hand, Symfind tends to follow subhalos with stable core particles much longer than Rockstar; as a result, core-collapse SIDM subhalos with tightly-bound central regions are recovered more accurately by Symfind. In velocity-dependent SIDM models invoked to explain recent small-scale structure anomalies, massive subhalos are generally core-expansion, and a rising fraction of lower-mass subhalos undergo core collapse~\cite{Nadler:2025jwh}. Thus, even though Symfind can identify more subhalos at small radii than Rockstar, the mass dependence of this improvement is non-trivial in SIDM, and in some cases, Symfind underperforms Rockstar.

As a result, tracking SIDM subhalos is more subtle and complex than in the CDM case, and the preferred subhalo finder can vary depending on factors such as particle count, tidal evolution history, and the gravothermal state of a given subhalo. We identify several general trends. For massive core-expansion subhalos and core-collapse subhalos that undergo close or repeated pericentric passages, a significant fraction of core particles can be lost, and Symfind does not provide a clear advantage. In contrast, for subhalos with large pericentric distances or fewer, more distant passages, Symfind tends to outperform Rockstar. Overall, for SIDM models with smaller cross section amplitudes and lower turnover velocities, Symfind’s performance improves.

Thus, SIDM physics can influence the relative performance of subhalo finders in a model-dependent fashion, which has important implications for SIDM analyses. For example, particle-tracking-based methods can underpredict the abundance of bright satellite galaxies by missing stripped, core-expansion subhalos, but they generally identify more core-collapse subhalos than phase-space methods, which is key for interpreting ultra-faint dwarfs and lensing perturbers. Given these complex tradeoffs, we ultimately recommend using a combination of \emph{both} algorithms to track SIDM subhalos and construct the most robust SIDM subhalo catalogs. In our subhalo mass function tests for two host halos with masses $10^{12}~{\rm M_\odot}$ and $10^{13}~{\rm M_\odot}$, the combined catalogs contain $2\textup{--}15\%$ more subhalos than either method alone when considering the overall population, and $10\textup{--}30\%$ more subhalos in the inner region of the host halo. For this reason, the subhalo catalogs in~\textsc{SIDM Concerto}~\cite{Nadler:2025jwh} are constructed using both methods.

We structure the rest of the paper as follows. We describe the simulations, subhalo finding algorithms, and our method to analyze the particle membership for individual subhalos in section~\ref{sec:method}. We conduct case studies on SIDM subhalos which are lost by halo finders in section~\ref{sec:subhalocore}, showing that a significant fraction of core particles become unbound from SIDM subhalos during tidal evolution. We validate this process with field halos and with idealized simulations. Then, we compare the subhalo mass functions in section~\ref{sec:shmf}. Furthermore, we explore how subhalo infall time and mass loss affect subhalo tracking in section~\ref{sec:subppl}. We conclude in section~\ref{sec:con}. In appendix~\ref{appex:sub-rho}, we present density profiles for representative examples discussed in section~\ref{sec:subhalocore}. In appendix~\ref{appex:core-ideal}, we show the evolution of core particles in idealized N-body simulations.

\section{Simulation Data and Methods}
\label{sec:method}

\subsection{Simulation Overview}
\label{subsec:simov}

We take CDM and SIDM cosmological zoom-in simulations of a strong lens analog (Group) and a Milky Way (MW) analog from Refs.~\cite{2023ApJ...958L..39N} and~\cite{2023ApJ...949...67Y}, respectively. These simulations are included in the \textsc{SIDM Concerto} compilation~\cite{Nadler:2025jwh} as Halo352 (Group) and Halo004 (MW). Their CDM versions were originally developed as part of the Symphony zoom-in compilation~\cite{2023ApJ...945..159N, Mao:2015yua}, which we will refer to as Group-CDM and MW-CDM. The SIDM simulations assume a strong, velocity-dependent SIDM model, characterized by the differential cross section:

\begin{equation}
\label{eqn:sidmdiff}
\frac{d \sigma}{d \cos \theta}=\frac{\sigma_0 w^4}{2\left[w^2+v^2 \sin ^2(\theta / 2)\right]^2},
\end{equation}
where $\sigma_{0}$ is the overall normalization factor, $w$ is the parameter controlling the velocity dependence, and $\theta$ is the scattering angle in the center of mass frame. These parameters are related to the fundamental properties of SIDM particles; see ~\cite{2010PhLB..692...70I, 2010PhRvL.104o1301F, 2013PhRvD..87k5007T, 2021MNRAS.505.5327T, 2022JCAP...09..077Y} for details. 

The Group-SIDM simulation takes $\sigma_{0}/m = 147.1 ~\rm{cm^{2}\,g^{-1}}$ and $w=120~\rm{km\,s^{-1}}$. In~\cite{2023ApJ...958L..39N}, this model has been shown to explain the extreme density diversity spanning from the dense perturber of the strong lensing system SDSSJ0946+1006~\cite{Vegetti:2009cz,Minor:2020hic} to the diffuse halos of gas-rich ultra-diffuse galaxies~\cite{PinaMancera:2021wpc,ManceraPina:2020ujo}. For both CDM and SIDM Group simulations, the particle mass is $m_{p} \approx 2.82 \times 10^{5} ~\rm{M_{\odot} \,h^{-1}}$ and the Plummer-equivalent gravitational softening length is $\epsilon = 170 ~\rm{pc\,h^{-1}}$; see~\cite{Nadler:2025jwh} for the cosmological parameters adopted in the simulations. Additionally, for comparison, we will show another group SIDM simulation with $\sigma_{0}/m = 70 ~\rm{cm^{2}\,g^{-1}}$ and $w=120~\rm{km\,s^{-1}}$ when discussing the subhalo mass functions in section~\ref{sec:shmf}. This simulation (Group-SIDM70) is based on Halo352 as well and included in the \textsc{SIDM Concerto} compilation~\cite{Nadler:2025jwh}.

The MW-SIDM simulation adopts $\sigma_{0}/m = 147.1 ~\rm{cm^{2}\,g^{-1}}$ and $w=24.33~\rm{km\,s^{-1}}$ (MW-MilkyWaySIDM), which are motivated by the need to explain the diverse DM distributions in MW satellite galaxies~\cite{2023ApJ...949...67Y}. Both MW-CDM and -SIDM simulations use $m_{p} \approx 4 \times 10^{4} ~\rm{M_{\odot}\,h^{-1}}$ and $\epsilon = 80 ~\rm{pc\,h^{-1}}$. In addition, the MW analog has also been simulated with the GroupSIDM model $\sigma_{0}/m = 147.1 ~\rm{cm^{2}\,g^{-1}}$ and $w=120~\rm{km\,s^{-1}}$~\cite{Nadler:2025jwh}. We refer to this run as MW-GroupSIDM in section~\ref{sec:shmf} when presenting the subhalo mass functions. For a comparison of the self-interaction cross section as a function of velocity for these simulated SIDM models, see figure 1 of~\cite{Nadler:2025jwh}.

For explicit demonstrations of the evolution of individual subhalos, we focus on those from the Group analog, which contains significantly more subhalos than the MW analog, making it easier to identify and compare matched CDM and SIDM subhalo pairs. Additionally, the Group-SIDM simulation features both a high self-interaction cross-section amplitude $\sigma_0/m$ and a high velocity turnover scale $w$, making its subhalos ideal for illustrating the novel challenges associated with tracking SIDM subhalos. For these reasons, we pick representative examples from the Group-CDM and -SIDM simulations for case studies in section~\ref{sec:subhalocore}.

\subsection{Subhalo Finders and Catalogs}
\label{subsec:subfinder}

We apply the particle-tracking-based finder Symfind~\cite{2024ApJ...970..178M} to the simulations discussed in section~\ref{subsec:simov}. The {Symfind} algorithm tags $N=32$ most-bound core particles within a subhalo at the snapshot of infall, and traces them forward to future snapshots.
The descendant of that subhalo is determined by which density peaks in the tracked subhalo particles contain the most core particles in subsequent snapshots. For details about the Symfind algorithm, we refer the reader to~\cite{2024ApJ...970..178M}.

We compare the Symfind results to the post-processed catalogs from the Rockstar subhalo finder~\citep{2013ApJ...762..109B} and the Consistent-Trees merger tree code pipeline \citep{2013ApJ...763...18B}. { The catalog post-processing is also performed by Symfind, as described in Appendix A of~\cite{2024ApJ...970..178M}. We note that this post-processing step is key to enabling one-to-one subhalo matching with Rockstar catalogs. In particular, this step takes results from Rockstar-plus-Consistent-Trees as the initial input and re-processes the merger tree by removing false subhalos, unphysical mass accretion histories, and incorrect subhalo property measurements caused by Rockstar losing track of halos as they disrupt. The current version of Symfind focuses on all (sub)halos that ever orbit within $R_{\rm vir}$ of the host system. Thus, Symfind re-indexes (sub)halos based on their largest pre-infall mass, using information from the post-processed Rockstar catalog, thereby aligning Rockstar and Symfind catalogs. As we will show later, unique halo IDs from the Rockstar catalog assigned at simulation time are converted into Symfind indices for easier access (e.g., a subhalo with index $10$ at snapshot $z=2$ is the primary progenitor of the subhalo with index $10$ at snapshot $z=0$).}

{On the other hand, the Symfind algorithm only begins operating at the first snapshot in which a (sub)halo enters within $R_{\rm vir}$ of the host system; it then tracks the subhalo through to the end of the simulation. As we show in the next section, the Symfind catalogs contain information only after first infall, while adopting all information from Rockstar catalogs beforehand. Consequently, cross-comparisons between Rockstar and Symfind catalogs are one-to-one for the same subhalo (e.g., a subhalo with index $10$ in the post-processed Rockstar catalog corresponds to the subhalo with index $10$ in the Symfind catalog). We therefore use subhalo indices to combine Symfind and Rockstar catalogs without double-counting any subhalos. In cases where both Symfind and Rockstar identify a given subhalo, we adopt the subhalo properties measured by Rockstar.}

Furthermore, for each of the CDM subhalos, we identify its SIDM counterpart to allow for one-to-one comparisons between their evolution and pre-infall trajectories. Following~\cite{2023ApJ...949...67Y, 2023ApJ...958L..39N}, we take the same definition of the virial mass as $\Delta_{\rm vir} \approx 99$ times the critical density of the Universe at $z=0$ for the cosmological parameters adopted in the simulations \cite{Bryan:1997dn}. We also adopt the same definition for the peak mass $M_{\rm peak}$ as in~\cite{2024ApJ...970..178M}, i.e., the maximum mass value before a subhalo ever became a subhalo of a larger system. This prevents numerical spikes in mass found in subhalo finders like Rockstar or Subfind from biasing $M_{\rm peak}$. 
For simplicity, we will use ``RCT'' to denote the results from the Rockstar+Consistent-Trees pipeline for the rest of the paper.

\subsection{Validation of Particle Membership}
\label{subsec:partmem}

The key assumption of a particle-tracking-based subhalo finder, like Symfind, is that most of the tagged core particles will belong to the same subhalo throughout tidal evolution. To verify consistency between the Symfind and RCT results, we validate the particle membership of a subhalo in each snapshot by explicitly requiring the total energy (i.e., gravitational potential energy plus kinetic energy) to be negative. We adopt the position and velocity of a subhalo from the RCT catalog, or from the Symfind catalog if the subhalo is not present in the former. We take all particles within the virial radius $R_{\rm vir}$ of the subhalo position or three times the tidal radius $3R_{\rm tidal}$ if the virial radius is unavailable in the catalog. The reason for choosing such a large radius during the initial trial is to capture the situation when the most-bound particles diffuse out of the subhalo due to numerical artifacts or SIDM scatterings. Then, we perform the same iterative unbinding method as in~\cite{2024ApJ...970..178M} to remove any unbound particles. We note that our validation method only utilizes halo radius, velocity, and position data from the RCT and Symfind catalogs, without requiring particle membership information from these catalogs. 

Our method allows the host halo particles to be captured by subhalos. In our test, up to $5\%$ of subhalo particles can originally belong to the host halo when the subhalo traverses the outskirts of the host halo. As discussed in~\cite{2024MNRAS.533.3811D}, the boundedness determination is subject to the choice of numerical parameters. Our parameter choice is not unique and relies on the accuracy of the catalogs; thus, it is fragile at the pericenter of any subhalo orbit as the chosen particle groups are subject to the largest tidal force leading to no bounded particles after iterative unbinding. 

\subsection{Tracking Core Particles}
\label{subsec:sidmcores}

The Symfind algorithm relies on tagged core particles of each subhalo ($N=32$). This is a design choice to prevent tracking spurious density peaks, and especially the relics of tidally disrupted subhalos, as the subhalo descendant~\cite{2024ApJ...970..178M}. Although particle tracking significantly reduces ambiguity over the evolution of subhalos compared to overdensity-based halo finders, there are still cases where it is difficult for a particle-tracking subhalo finder to identify the density peak that corresponds to a given subhalo. For example, a long-lived subhalo may contain many orders of magnitude more mass in its tidal tails than in its bound core, and those tails can phase-mix into complex geometries with unexpected density peaks. Because of this, it is essential to identify only the density peak that corresponds to the subhalo. Similarly, in some cases, massive subhalos can truly merge with their hosts in cataclysmic ``major'' mergers, and the algorithm should not attempt to place them in some minor secondary density peak after the main peak has been subsumed. The use of core particles is Symfind's method for resolving this ambiguity. A similar approach is used by the subhalo finder Bloodhound~\citep{Kong_2025}. The main benefit of this method over an alternative approach~\citep[e.g.][]{2012MNRAS.427.2437H,2018MNRAS.474..604H,2024MNRAS.533.3811D}, where the algorithm gradually restricts the pool of tracked particles, is that it is insensitive to errors in previous snapshots. However, this benefit comes at the cost of not updating the core particles over time, meaning that Symfind is potentially more sensitive to physical and numerical effects that scatter particles out of the halo's center.

Therefore, the drawback of this approach is that the evolution of the core particles is very important to the success of the halo finder. If numerical or physical forces eject those particles from the centers of their subhalos, Symfind can have difficulty tracking them. Particularly, in the case of excessive mass loss and extreme tidal shocking, Symfind may identify the wrong density peaks as a descendant, or it may not find any descendant. Fortunately, in CDM, most of the core particles are typically bounded within the subhalo over its entire tidal evolution history, and Symfind is able to recover from temporary errors. Additionally, by the time numerical effects remove the core particles from the subhalo, the subhalo is generally so relaxed that it is not usable for analysis anyway~\cite{2024ApJ...970..178M}. However, this is not generally the case in SIDM, where the combination of self-scattering and tidal stripping effects causes core particles to be ejected from the subhalo more easily, as we show next.

\section{Evolution of Core Particles in CDM and SIDM Subhalos}
\label{sec:subhalocore}

\subsection{Core-expansion SIDM Subhalos}
\label{subsec:coreform}

\begin{figure*}[h!]
  \centering
    \includegraphics[width=\columnwidth]{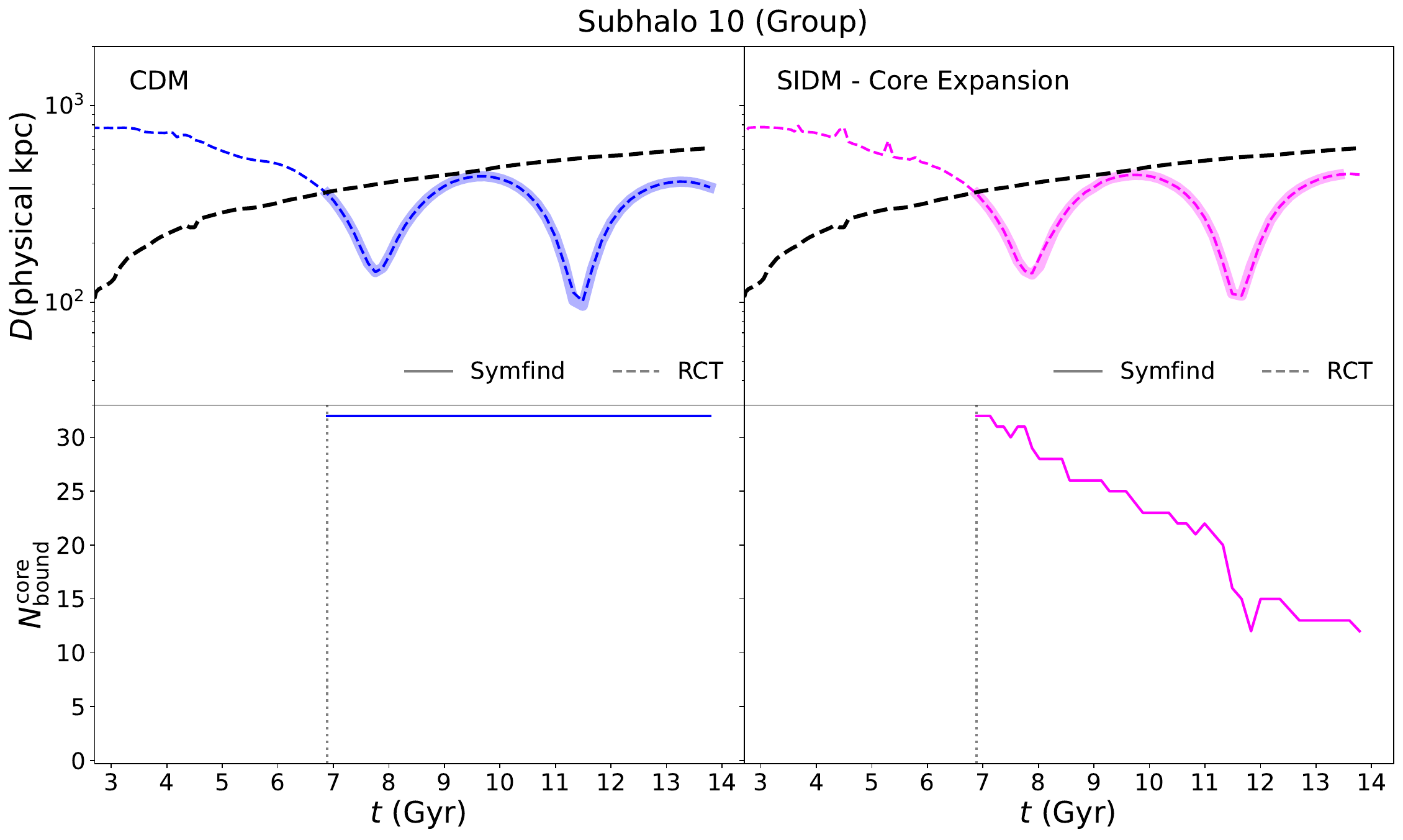}
    \caption{The top row shows the orbital tracks of Subhalo~10, which has a peak mass of $M_{\rm peak} = 1.3\times10^{11}~\rm{M_{\odot}}$, from the Group-CDM (left, blue) and -SIDM (right, magenta) simulations. We plot the distance between the subhalo and the center of the host halo as recorded in the Symfind catalog (solid) and the RCT catalog (dashed). The dashed black line indicates the virial radius of the host halo. The bottom row shows the number of core particles that remain bound to the subhalo as a function of time. The dotted gray line marks the infall time.}
    \label{fig:group-case}
\end{figure*}

In figure~\ref{fig:group-case}, we show the tidal evolution of Subhalo~10 from the Group-CDM (left) and -SIDM (right) simulations. Its peak mass is $M_{\rm peak} = 1.3\times10^{11}~\rm{M_{\odot}}$ and this is one of the most massive subhalos. In the top left panel, we show the distance between the CDM subhalo and the host halo center as a function of time, generated from the Symfind catalog (solid blue) and the RCT catalog (dashed blue). For reference, we also plot the viral radius of the host halo (dashed black). In the bottom left panel, we plot the number of core particles that are still bound to the subhalo. We see that all core particles, which are initially tagged, are bound to the CDM subhalo for the entire tidal evolution history, and both Symfind and RCT methods track this subhalo to $z=0$.

In figure~\ref{fig:group-case}, we also show the SIDM counterpart of Subhalo~10. As indicated in the top right panel, Symfind regards that the SIDM subhalo is disrupted at the last apocenter passage at $t=13.4~\rm{Gyr}$ (solid magenta), while RCT keeps tracking it (dashed magenta). The core particles of the SIDM subhalo become continuously less bound, and the loss is more severe after each pericenter passage, as shown in the bottom right panel. When Symfind loses track of this subhalo at $t=13.4~\rm{Gyr}$, there are still $10$ core particles bound to the subhalo. Thus, this tracking failure is not caused by losing all core particles in the subhalo, but rather due to the fact that Symfind could not identify the correct descendant when the majority of the core particles are lost.

\begin{figure*}[!]
  \centering
    \includegraphics[width=\columnwidth]{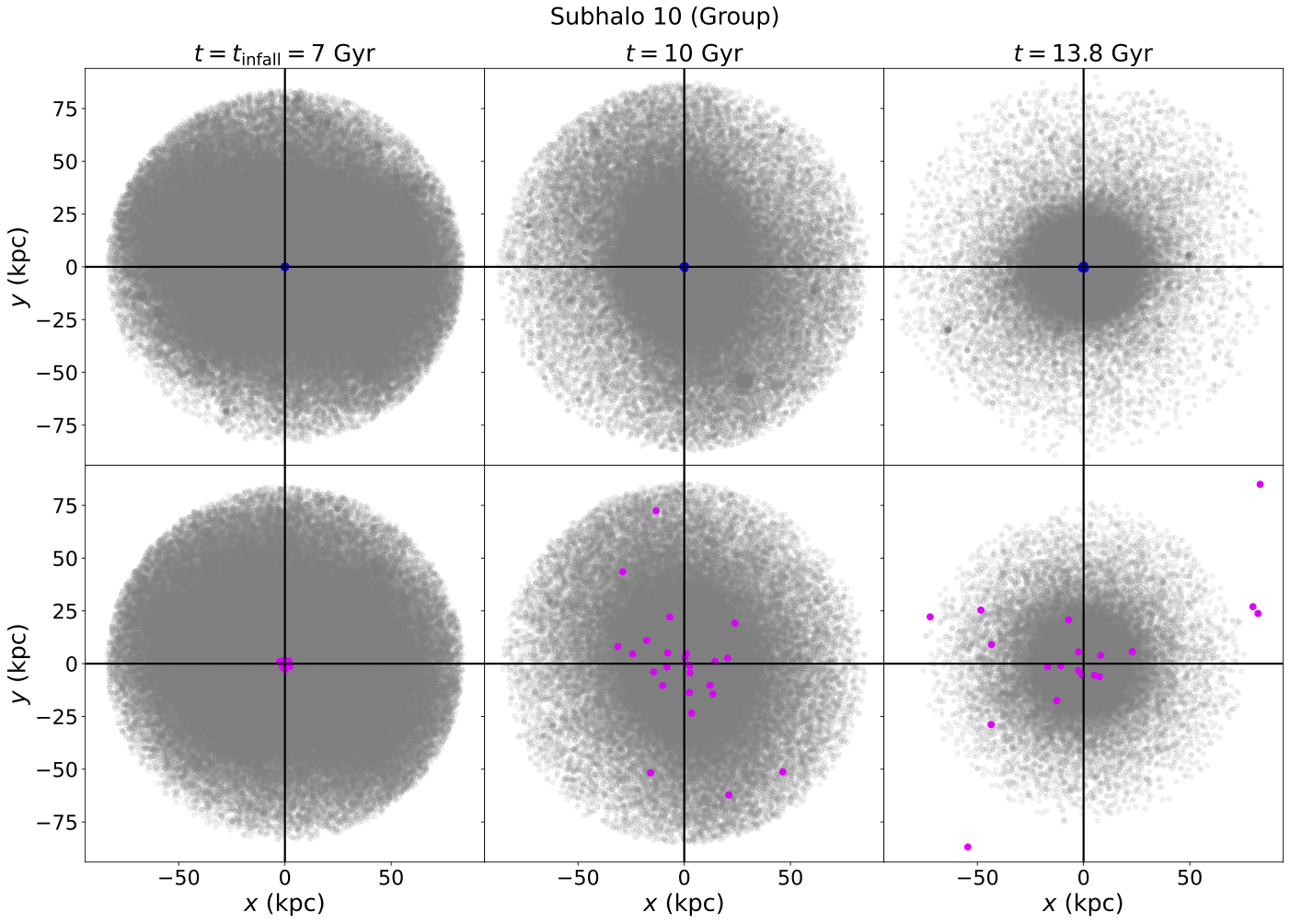}\\
    \includegraphics[width=\columnwidth]{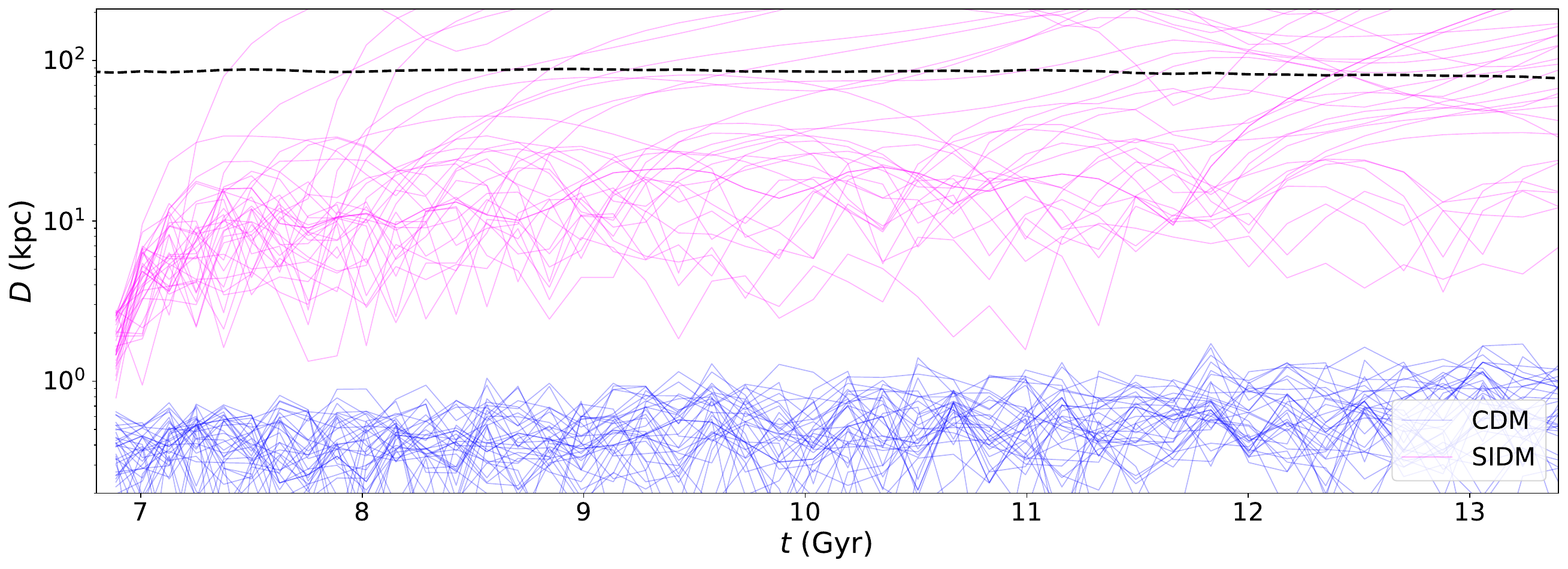}
    \caption{The top and middle panels show the positions of bound particles in Subhalo~10 from the Group-CDM and -SIDM simulations, respectively. The positions are given in physical units and projected onto the $x\textup{--}y$ plane at three different times, centered on the subhalo. Gray dots represent non-core particles, while colored dots indicate core particles. The bottom panel shows the distance of core particles from the subhalo center as a function of time. The dashed black line marks the virial radius of the SIDM subhalo, which serves as the maximum boundary of the subhalo in this analysis. Blue and magenta indicate the CDM and SIDM simulations, respectively.}
    \label{fig:group-pos}
\end{figure*}

In the top and middle panels of figure~\ref{fig:group-pos}, we show the particle position with respect to the subhalo center for the CDM subhalo and its SIDM counterpart, respectively, at $t=t_{\rm infall}=7~\rm{Gyr}$, $10~{\rm Gyr}$, and $13.8~{\rm Gyr}$. All bound particles are shown, while the core particles are highlighted in magenta and blue for SIDM and CDM, respectively. The core particles of the CDM subhalo remain at the center for the entire time shown. In contrast, although the core particles of the SIDM subhalo are located at the center at infall, many of them are kicked out and then stripped at later snapshots. 

In the bottom panel of figure~\ref{fig:group-pos}, we show the evolution of the distance from the core particles to the center of the subhalo. For the CDM subhalo (blue), there is a very mild overall trend that causes the core particles to diffuse outward over time. This is likely caused by numerical two-body relaxation in the halo's center~\citep[e.g.][]{Power:2003,Ludlow:2019}. In contrast, the SIDM counterpart's core particles migrate to much larger distances over time due to a combination of self-interaction and tidal stripping. After tidal stripping, some can be found hundreds of kpc away from the center of the subhalo. 

The SIDM counterpart of Subhalo~10 has a $\sim10~{\rm kpc}$ density core as shown in appendix~\ref{appex:sub-rho}. We note that SIDM subhalos with large cores are more prone to the issues of tracking core particles. A key factor is that the core particles may migrate beyond the core radius, as we show in the next section. For a massive subhalo, initially, the outer region can act as a buffer to ``catch'' the core particles once they are kicked out beyond the core radius. As tidal evolution begins, mass is removed from a subhalo's outer region, but the inner region is undisturbed at first; thus, the core remains intact. After sufficient tidal stripping, most of the outer region gets removed while the subhalo is left with an isothermal core. At this stage, core particles are rapidly removed from the subhalo by a combination of DM self-interactions and tidal forces, leading to an incorrect disruption determination with the particle-tracking-based method. 

\subsection{Evolution of Core Particles in Isolated Halos}
\label{subsec:isolatedcore}

\begin{figure*}[h!]
    \centering
    \includegraphics[width=\columnwidth]{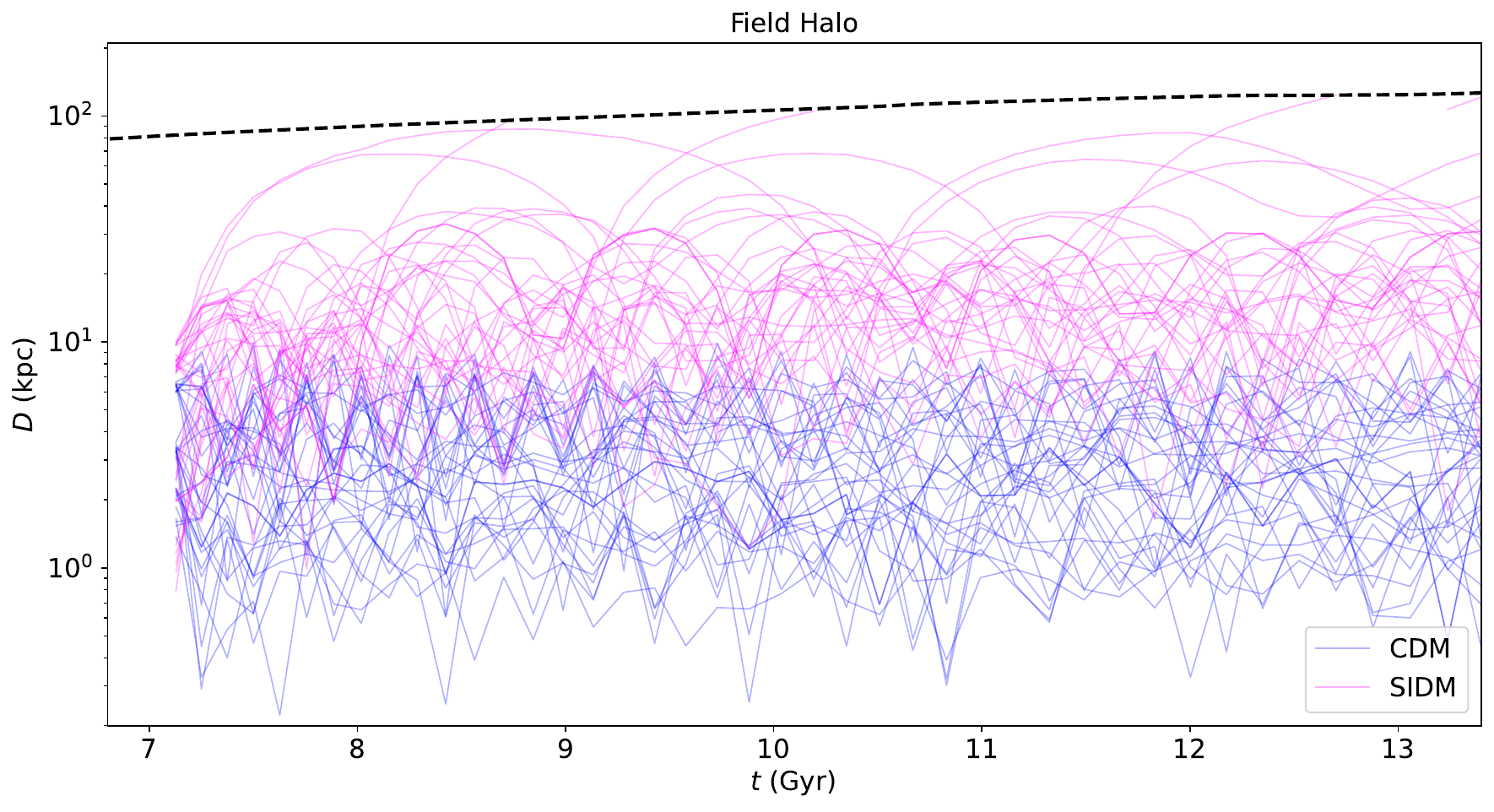} 
    \caption{The distance of core particles, tagged at $7~{\rm Gyr}$, from the halo center as a function of time for a pair of field halos in the Group-CDM (blue) and -SIDM (magenta) simulations. The dashed black line marks the virial radius of the SIDM halo.}
    \label{fig:iso-pos}
\end{figure*}

To further explore the effects of SIDM on core particles, we analyze field halos from the Group-CDM and -SIDM simulations to isolate the impact of self-interactions versus tidal stripping. We pick a matched pair of CDM and SIDM halos from the extended high-resolution regions around the main hosts. Their masses are $\sim10^{11} ~\rm{M_{\odot}}$ at $z=0$ and they do not experience major mergers after $7~\rm{Gyr}$, which is the infall time of the subhalo discussed in section~\ref{subsec:coreform}. The distance of the field halos to their corresponding primary host halos is $\gtrsim1~\rm{Mpc}$ at any given snapshot. We tag their core particles at $t\approx 7~\rm{Gyr}$, by which time they have assembled $85\%$ of their final masses. In figure~\ref{fig:iso-pos}, we show the evolution of the distance of the core particle from the center for the CDM (blue) and SIDM (magenta) field halos. For the CDM halo, the core particles stay within $\sim7~{\rm kpc}$ and have stable orbits. In contrast, for the SIDM halo, the core particles can migrate to much larger distances, with apocenters close to $\approx30~{\rm kpc}$, and a few of them can reach $80~{\rm kpc}$. 

We have further conducted idealized simulations of isolated SIDM and CDM halos with a similar mass $ 10^{11} ~\rm{M_{\odot}}$ and present them in appendix~\ref{appex:core-ideal}. In the absence of a cosmological environment, the SIDM core particles still migrate from the inner region to the outer region of the halo. Such behavior is the direct consequence of the DM self-interactions. In particular, collisions transfer momentum between DM particles, increasing the kinetic energy of core particles and moving them to orbits with a larger radius.

\subsection{Tracking core-collapse SIDM Subhalos}
\label{subsec:trackcc}

So far, we have focused on core-expansion massive SIDM (sub)halos. These systems are prone to the loss of core particles, and as a result, a particle-tracking-based subhalo finder such as Symfind does not always offer a clear advantage in tracking them. In this section, we turn our attention to core-collapse subhalos and investigate the factors that influence the performance of subhalo finders. In this regime, both Symfind and RCT can provide complementary and valuable information for subhalo identification and tracking.

\begin{figure*}[!]
  \centering
    \includegraphics[width=\columnwidth]{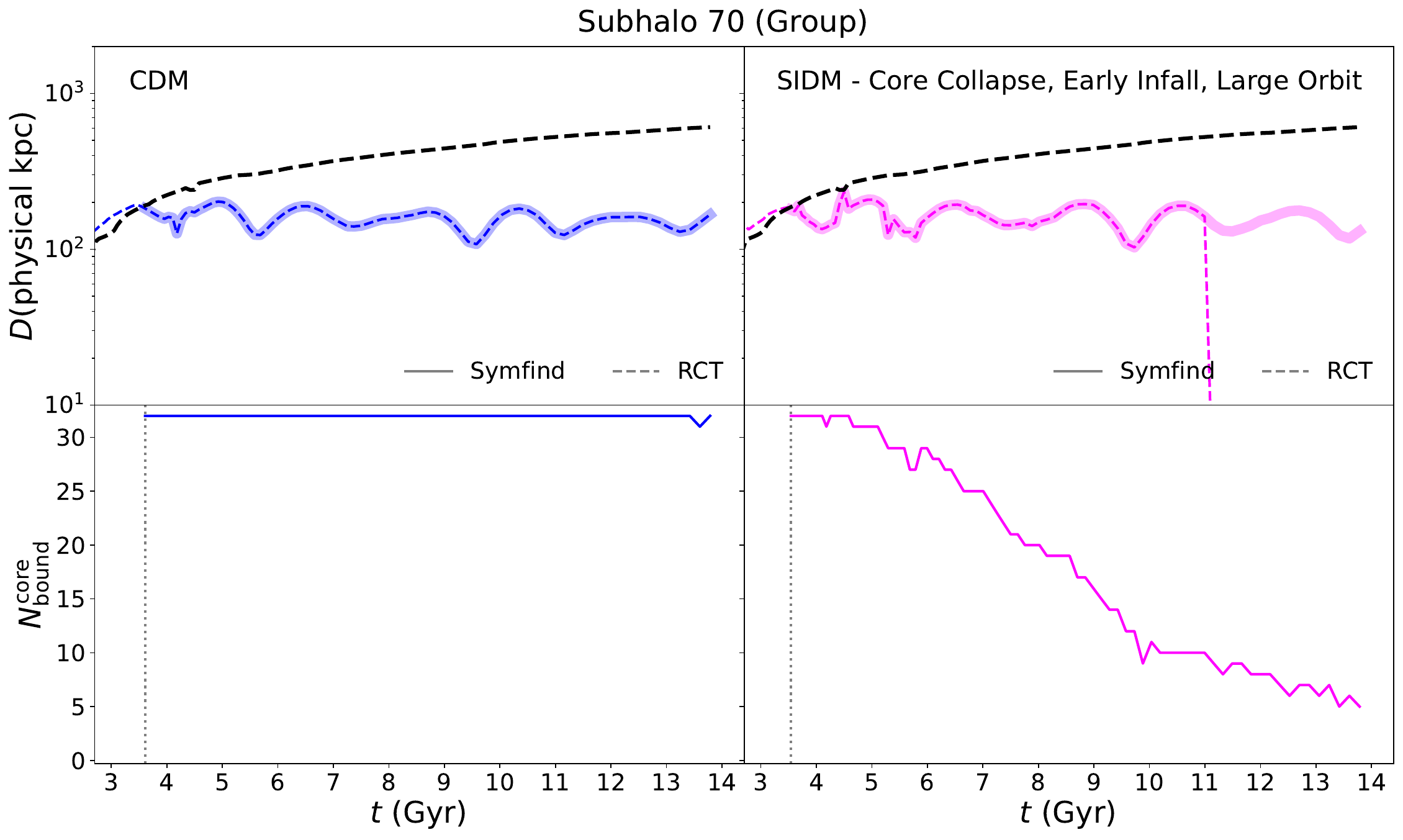}
    \caption{Similar to figure~\ref{fig:group-case}, but for Subhalo~70 with $M_{\rm peak} = 1\times10^{10}~\rm{M_{\odot}}$ from the Group simulations. Note that ``70" refers to the subhalo index in the Group-SIDM simulation, which corresponds to Subhalo~139 in the Group-CDM simulation. In the SIDM case, the RCT method loses the subhalo around $11~\rm{Gyr}$.}
    \label{fig:group-cc-lost-case-70}
\end{figure*}

\begin{figure*}[!]
  \centering
    \includegraphics[width=\columnwidth]{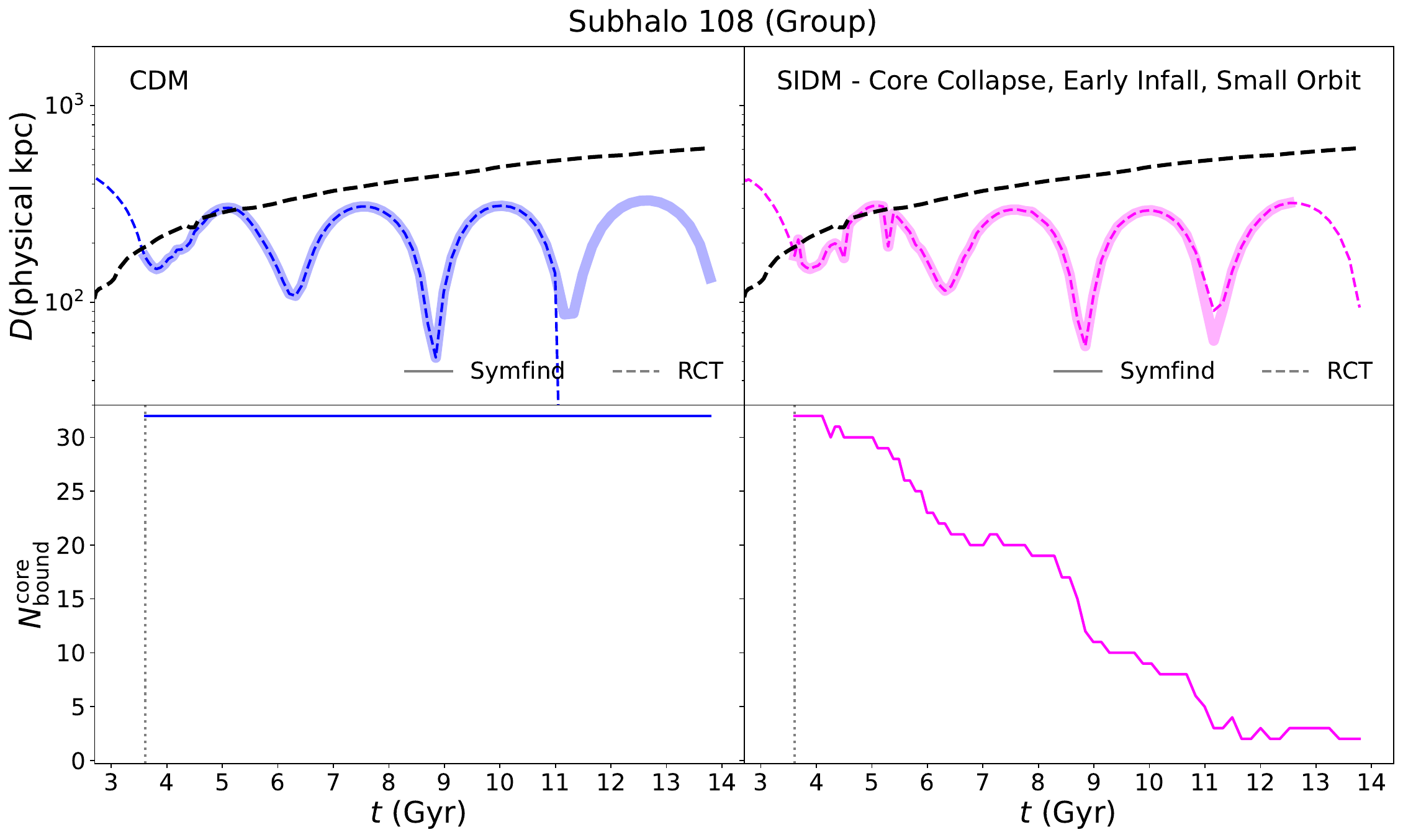}
    \caption{Similar to figure~\ref{fig:group-case}, but for Subhalo~108 with $M_{\rm peak} = 6.8\times10^{9}~\rm{M_{\odot}}$ from the Group simulations. Note that ``108" refers to the subhalo index in the Group-SIDM simulation, which corresponds to Subhalo~124 in the Group-CDM simulation. In the CDM case, the RCT method loses track of the subhalo around $11~\rm{Gyr}$, while in the SIDM case, the Symfind method loses the subhalo around $12.5~\rm{Gyr}$. }
    \label{fig:group-cc-lost-case-108}
\end{figure*}

\begin{figure*}[!]
  \centering
    \includegraphics[width=\columnwidth]{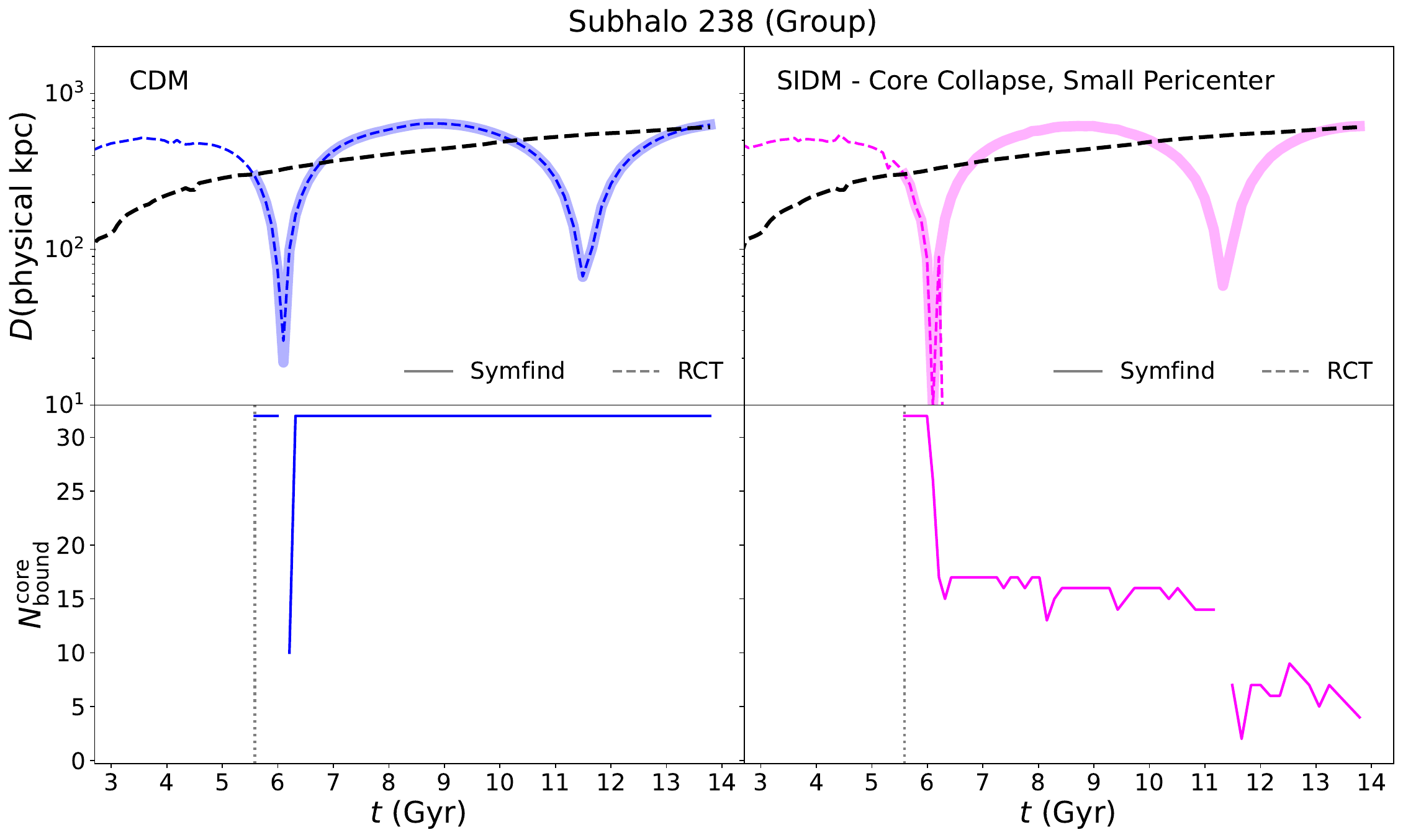}
    \caption{Similar to figure~\ref{fig:group-case}, but for Subhalo~238 with $M_{\rm peak} = 2.6\times10^{9}~\rm{M_{\odot}}$ from the Group simulations. Note that ``238" refers to the subhalo index in the Group-SIDM simulation, which corresponds to Subhalo~281 in the Group-CDM simulation. In the SIDM case, the RCT method loses track of the subhalo around $6~\rm{Gyr}$.}
    \label{fig:group-cc-lost-case-238}
\end{figure*}

In figure~\ref{fig:group-cc-lost-case-70}, we present Subhalo~70 which has a very early infall at $3.5~\rm{Gyr}$ with a peak mass $M_{\rm peak} = 1\times10^{10}~\rm{M_{\odot}}$. For its CDM counterpart, both Symfind and RCT can keep track of it. For comparison, the RCT method lost track of the SIDM subhalo around $11~\rm{Gyr}$, while Symfind can still keep track of it in the situation of continuous particle loss. We note that this subhalo has a large orbital radius around $100~\rm{kpc}$, thus the core particles are lost in a relatively stable speed due to weaker tidal interaction, causing no trouble to the particle-tracking-based method.

In figure~\ref{fig:group-cc-lost-case-108}, we show Subhalo~108 with $M_{\rm peak} = 6.8\times10^{9}~\rm{M_{\odot}}$. This subhalo has a similar infall time to Subhalo~70, but its pericenter, around $50~\rm{kpc}$, is smaller. For its CDM counterpart, this orbital radius and long tidal interaction time cause the RCT to drop it at the last pericenter passing around $11~\rm{Gyr}$. As the core particles are stable within the CDM subhalo, the Symfind has no problem tracking this CDM subhalo until the end of the simulation. On the other hand, Symfind drops the SIDM subhalos at the last apocenter passing due to excessive loss of core particles. Compared to Subhalo~70, the core particles in Subhalo~108 can be lost relatively quickly due to closer pericenter passing, suggesting that a combination of long tidal evolution times and close orbital radii can lead to a non-tracking problem for the particle-tracking-based method.

In figure~\ref{fig:group-cc-lost-case-238}, we present Subhalo~238 with $M_{\rm peak} = 2.6\times10^{9}~\rm{M_{\odot}}$, illustrating a case of a close pericenter passage with a shorter tidal evolution time. Compared to its CDM counterpart, the SIDM subhalo has a closer pericenter passing less than $10~\rm{kpc}$, which causes the subhalo to be lost in the RCT method. In comparison, although the SIDM subhalo lost almost half of its core particles in this pericenter passing, Symfind is still able to continue tracking this subhalo until the end of the simulation. Note that during the first pericenter in the CDM simulation and the second pericenter in the SIDM simulation, Symfind temporarily loses track of the subhalo, but is able to recover it again at future times. We also present the density profiles of Subhalos~70, 108, and 238 in appendix~\ref{appex:sub-rho} for reference.

\subsection{Summary of Core Particle Tracking Results}

We have shown that core particles in SIDM subhalos are less likely to stay bound within the central regions compared to CDM subhalos. In particular, DM self-interactions can cause core particles to migrate to the outer region of the subhalo, and tidal forces can subsequently unbind them from the system. As a result, the tradeoff between particle-tracking-based (Symfind) versus phase-space-based (RCT) subhalo finders depends on both gravothermal evolution stage and tidal orbit of a given system as follows. 

\begin{itemize}
  \item For all types of SIDM subhalos, a close pericenter passage leads to excessive loss of core particles. 
  \item For the most massive subhalos (Subhalo~10), SIDM physics typically produces large cores in the inner regions for SIDM models with a large cross section. Once the NFW-like region of an SIDM subhalo is mostly stripped by tidal forces, core particles are more easily lost during subsequent pericenter passages, potentially leading to an incorrect determination of disruption. As a result, a particle-tracking-based subhalo finder, such as Symfind, does not always provide an advantage for SIDM subhalos with large inner cores.
  \item For core-collapse subhalos with large pericentric distances and early infall times (Subhalo~70), Symfind often outperforms RCT.
  \item For core-collapse subhalos with near and multiple pericentric passages~(Subhalo~108), the majority of core particles will be lost and cannot be reliably tracked by Symfind.
  \item For core-collapse subhalos with closer but fewer pericentric passages~(Subhalo~238), Symfind is preferred as the subhalos traverse the densest region of the host system, where RCT struggles to identify subhalos reliably.
\end{itemize}

\section{Subhalo Mass Functions}
\label{sec:shmf}

\begin{figure*}[h!]
  \centering
    \includegraphics[width=\columnwidth]{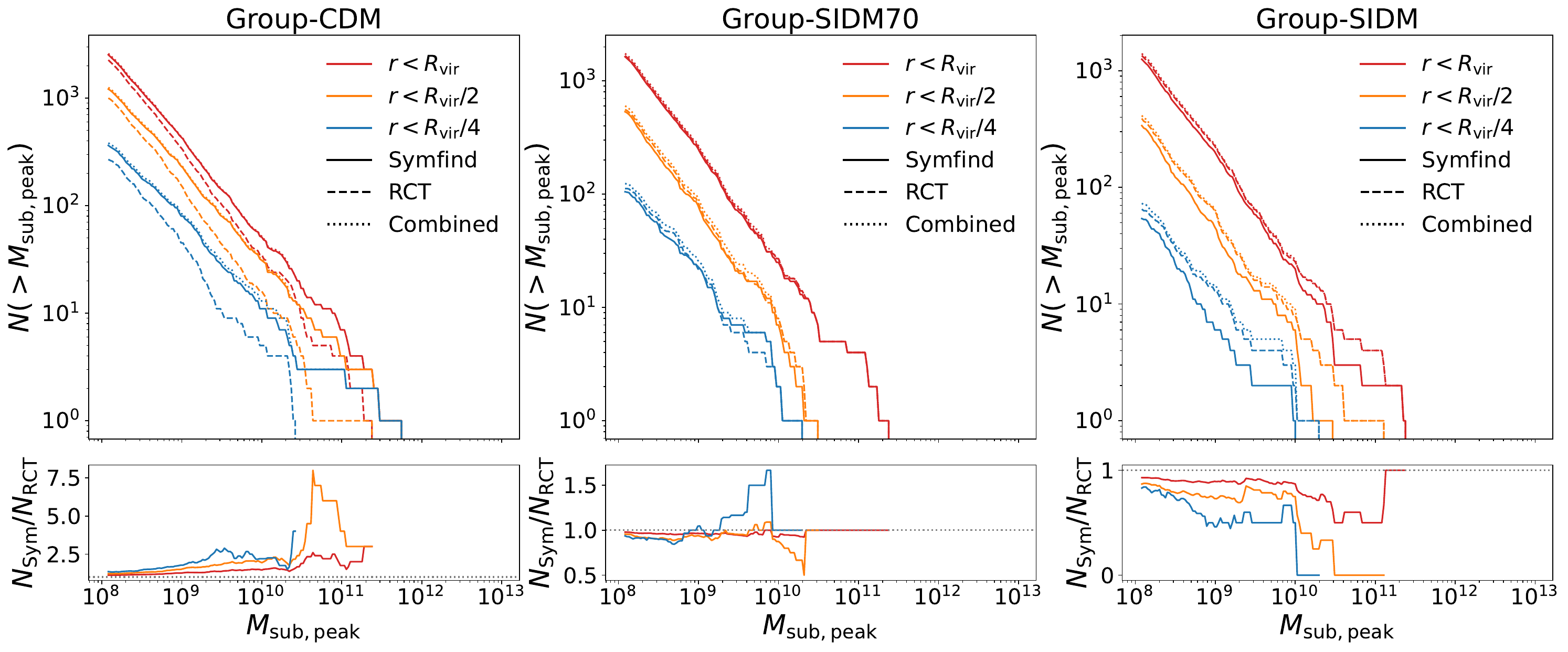}
    \caption{Subhalo $M_{\rm peak}$ functions at $z=0$ in the Group-CDM (left), -SIDM70 (middle), and -SIDM (right) simulations.
Top panels: Subhalo mass functions from the Symfind (solid), RCT (dashed), and combined (dotted) catalogs. Different colors represent results within different radial bins. Bottom panels: Ratios of the Symfind to RCT mass functions. The dotted horizontal line indicates a ratio of unity.}
    \label{fig:group-syf-rs}
\end{figure*}

\begin{figure*}[h!]
  \centering
    \includegraphics[width=\columnwidth]{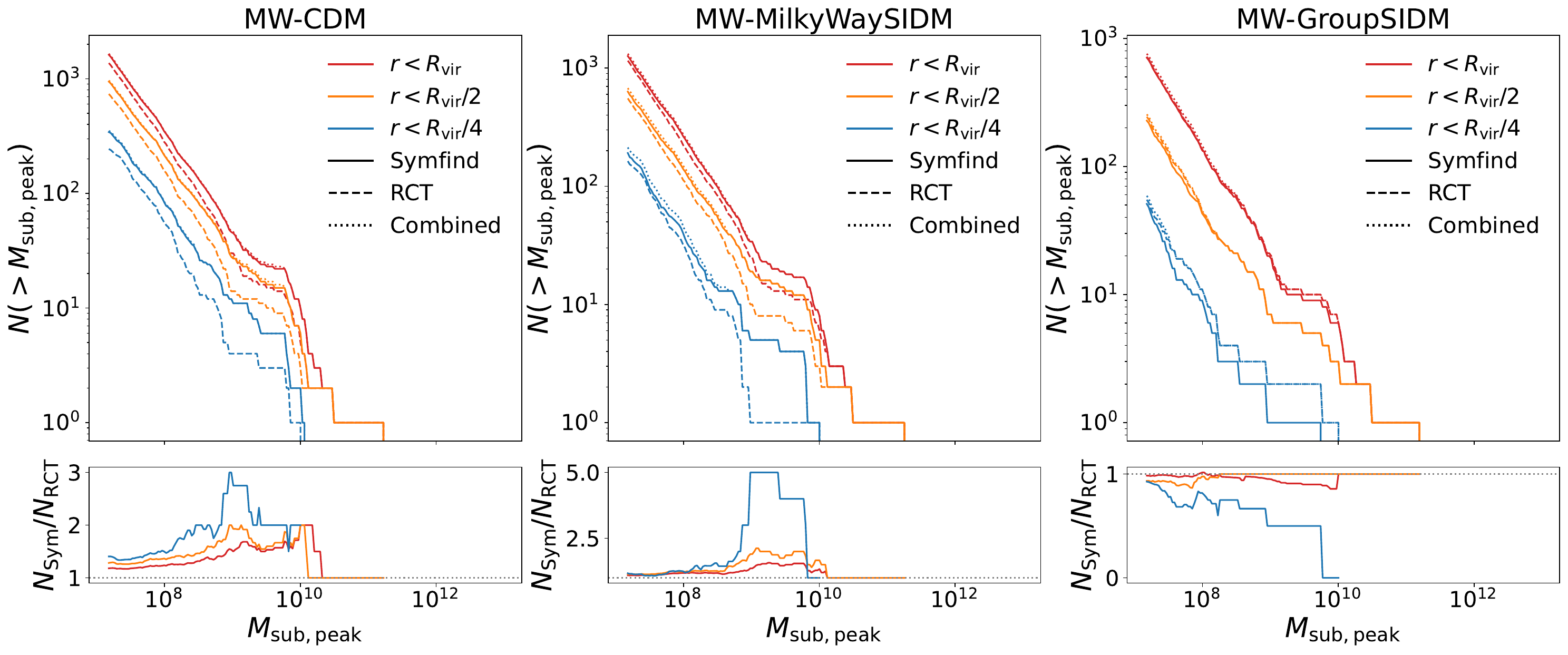}
    \caption{Similar to figure~\ref{fig:group-syf-rs}, but for the MW-CDM (left), -MilkyWaySIDM (middle), and -GroupSIDM (right) simulations.}
    \label{fig:mw-syf-rs}
\end{figure*}

With this intuition, we now study the entire subhalo populations in our Group and MW simulations, which contain a combination of all types of subhalos discussed above. In figure~\ref{fig:group-syf-rs}, we compare subhalo $M_{\rm peak}$ mass functions from the RCT and Symfind catalogs for the Group-CDM (left), -SIDM70 (middle), and -SIDM (right) simulations. For the subhalo peak mass functions from the Group-CDM simulation, Symfind systematically outperforms RCT across all radial bins, with the difference becoming more pronounced at smaller radii. Notably, Symfind extends the subhalo mass functions for the inner region ($r < R_{\rm vir}/2$) towards large mass scales ($M_{\rm peak} > 3 \times 10^{10}~\rm{M_{\odot}}$), where RCT does not recover any subhalos. For subhalos in the Group-SIDM70 simulation, where the amplitude of the cross section is $\sigma_0/m=70~{\rm cm^2\,g^{-1}}$ and the velocity turnover $w=120~{\rm km\,s^{-1}}$, the performances of Symfind and RCT become comparable overall. For those in the inner region ($r < R_{\rm vir}/4$), Symfind outperforms mildly. As the amplitude further increases to $\sigma_0/m=147.1~{\rm cm^2\,g^{-1}}$, while keeping $w$ fixed, RCT outperforms Symfind for the subhalos in the Group-SIDM simulation. Compared to the Group-CDM simulation, the differences between the RCT and Symfind catalogs are exactly the opposite in the Group-SIDM simulation, as the Symfind catalogs are missing the most massive subhalos that are recovered by RCT for the reasons discussed in section~\ref{sec:subhalocore}.

In figure~\ref{fig:mw-syf-rs}, we compare the RCT and Symfind catalog subhalo $M_{\rm peak}$ mass functions from our MW-CDM (left), -MilkyWaySIDM (middle), and -GroupSIDM (right) simulations. For CDM, Symfind recovers $3$ times more subhalos than RCT in the inner regions of the host ($r < R_{\rm vir}/4$) for systems with $M_{\rm peak}\sim10^{9-10}~\rm{M_{\odot}}$ at $z=0$. This performance gain is also evident in the MW-MilkyWaySIDM simulation, where Symfind substantially improves recovery for subhalos in the inner region compared to RCT. In contrast, for the MW-GroupSIDM simulation, Symfind does not outperform RCT, particularly in the inner regions ($r<R_{\rm vir}/4$). Such differences come directly from the difference in SIDM models. The MilkyWaySIDM model assumes $\sigma_0/m=147.1~{\rm cm^2\,g^{-1}}$ and $w=24.33~{\rm km\,s^{-1}}$, while the GroupSIDM model adopts the same value of $\sigma_0/m$, but a higher $w=120~{\rm km\,s^{-1}}$. With a larger turnover velocity $w$, massive subhalos have a higher SIDM cross section and the loss rate of core particles becomes higher in the MW-GroupSIDM simulation, where Symfind loses high-$M_{\rm peak}$ subhalos in the inner region ($r < R_{\rm vir}/4$). This is similar to the Group-SIDM simulation,  suggesting that the result is independent of both resolution and system. It is interesting to note that Symfind and RCT perform comparably in recovering the total subhalo population within the virial radius in the MW-GroupSIDM simulation, which contrasts with the results from the Group-SIDM simulation. This difference likely arises because, in the former, most subhalos are in the collapse phase, whereas many are still in the core-expansion phase in the latter.

For both Group-CDM and MW-CDM simulations, combining the RCT and Symfind catalogs yields only marginal gains, with the combined catalogs containing less than $2\%$ more subhalos than Symfind alone and about $10\%$ more than RCT. In contrast, for SIDM simulations, the combined approach offers a substantial improvement, recovering up to $30\%$ more subhalos, especially in the inner regions of the host halo. We therefore recommend combining Symfind and RCT results to construct the most complete and robust SIDM subhalo catalogs.

More specifically, for the Group simulations, a combined catalog provides overall $2\%$ more subhalos than the RCT catalog and $10\%$ more subhalos than the Symfind catalog. In the innermost region of $r<R_{\rm vir}/4$, the combined catalog contains $10\%$ more subhalos than the RCT catalog and $30\%$ more than the Symfind catalog. For the MW simulation with the MilkyWay SIDM model, the combined catalog includes $15\%$ more subhalos than the RCT catalog and $5\%$ more than the Symfind catalog overall. In the inner region of $r<R_{\rm vir}/4$, it provides $30\%$ more subhalos than the RCT catalog and $10\%$ more than the Symfind catalog. Similarly, for the MW simulation with the GroupSIDM model, the combined catalog contains $8\%$ more subhalos than both the RCT and Symfind catalogs overall. In the inner region of $r<R_{\rm vir}/4$, it recovers $8\%$ more subhalos than the RCT catalog and $15\%$ more than the Symfind catalog. Although the particle-tracking-based methods are affected by DM collision physics, they are still able to recover many subhalos missed by phase-space methods. The results from both approaches are thus complementary.

We have also checked that increasing the number of core particles from $N_{\rm core}=32$ to $128$ in Symfind does not improve the subhalo population recovery in the high-$M_{\rm peak}$ regime for the GroupSIDM model. { On the contrary, increasing the number of core particles leads to fewer recovered subhalos with peak particle counts} $N_{\rm peak} > 10^{3.5}$. This is likely because Symfind's method for determining subhalo position is based on finding the density peak that is associated with the largest fraction of core particles. Most cases of disruption are due to the true remnant containing relatively few core particles relative to the extended density peak represented by the tidal tails and extended phase-mixed mass component, and increasing core particle count is unlikely to help. Once the core particle count is large enough to prevent shot-count effects, increasing the particle count likely only dilutes the core particle pool with less bound particles.

Moreover, we have also tested whether redefining core particles at each snapshot can alleviate the challenges of tracking SIDM subhalos. In principle, one could identify the most bound particles at every snapshot, label them as core particles, and then follow their evolution. However, this approach requires accurate determinations of the positions and velocities of individual particles relative to the main descendant of the subhalo, which is often difficult to achieve in a tidal environment. In our tests, we found that simply reconstructing the core particles at each snapshot does not improve tracking success. 

It is also possible to track all particles of a subhalo, as implemented in algorithms such as those in~\cite{2018MNRAS.474..604H,2024MNRAS.533.3811D}, which have demonstrated good performance in the CDM case. Detailed tests are required to assess their effectiveness in SIDM scenarios. In cases of severe mass loss due to the combined effects of self-interactions and tidal stripping, we suspect that it may be unclear how to meaningfully define or include such subhalos in a catalog, beyond simply identifying and following all of their particles. We emphasize that the primary challenge in tracking SIDM subhalos arises from the difficulty of accurately reconstructing their positions and velocities from tracked particles and identifying the correct main descendant, rather than from an inability to follow particles over the course of tidal evolution. Therefore, relying solely on particle positions to determine subhalo descendant properties may be insufficient for SIDM simulations.

\section{Mass Ratio and Infall Time of Subhalos}
\label{sec:subppl}

\begin{figure*}[h!]
  \centering
    \includegraphics[width=\columnwidth]{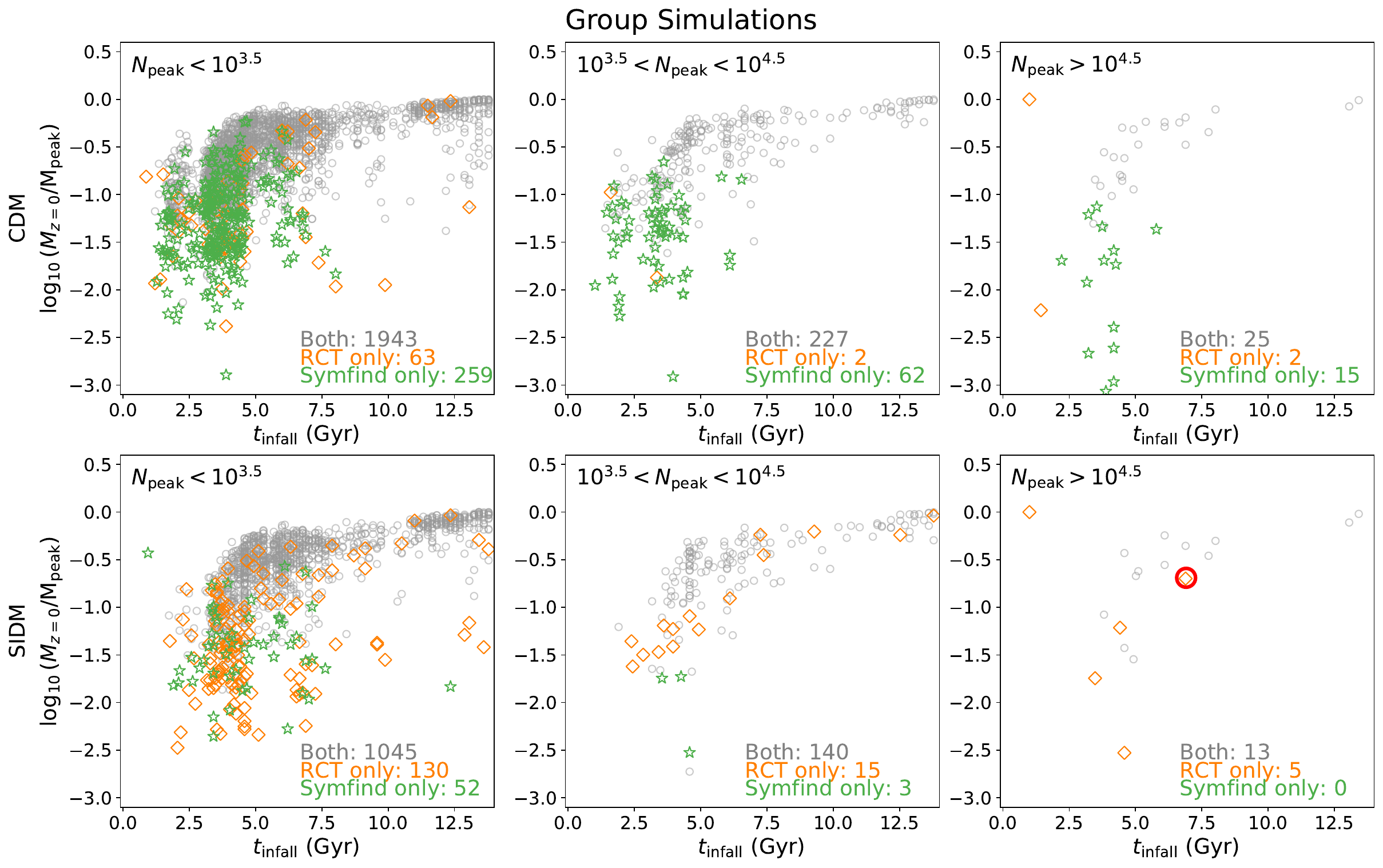}
    \caption{Ratio of the subhalo mass at $z=0$ to its peak mass $M_{\rm peak}$ versus infall time $t_{\rm infall}$ for subhalos in the Group-CDM (top) and -SIDM (bottom) simulations, shown in three bins of peak particle number $N_{\rm peak}$. $t_{\rm infall}$ denotes the time of subhalo first infall to the main host. Subhalos are categorized into three groups: gray circles denote those found in both the Symfind and RCT catalogs at $z=0$; orange diamonds denote those found only in the RCT catalog; and green stars denote those found only in the Symfind catalog. The number of subhalos in each group is indicated in the bottom-left corner of each panel. The red circle marks Subhalo~10, also examined in figure~\ref{fig:group-case}.}
    \label{fig:subov-group}
\end{figure*}

In this section, we investigate how infall time and mass loss affect the relative performance of Symfind and RCT. As in section~\ref{sec:subhalocore}, we again take the subhalos in the Group-CDM and -SIDM simulations for the case study. Figure~\ref{fig:subov-group} shows the ratio of the subhalo mass at $z=0$ to its peak mass $M_{\rm peak}$ versus infall time $t_{\rm infall}$. We take all subhalos from the Symfind and the RCT catalogs at $z=0$ and categorize them into three groups: subhalos found in both Symfind and RCT catalogs (gray circles); those only in the RCT catalog (orange diamonds); and those only in the Symfind catalog (green stars). Furthermore, we divide the subhalos into bins based on their peak particle count, $N_{\rm peak}$, to explore these differences as a function of peak subhalo mass. Note we define the infall time $t_{\rm infall}$ as the time of first infall to the main host. This does not necessarily mean that a subhalo spends all the later time within the virial radius of the host halo. In some cases, splashback subhalos can traverse the virial radius multiple times throughout the evolution and spend large amounts of time in very low density regions. 

In the top panels of figure~\ref{fig:subov-group}, we show CDM subhalos. Both Symfind and RCT find the majority of subhalos, while the disagreements account for $10\%$ to $20\%$ of the overall population. For small subhalos that have $N_{\rm peak} < 10^{3.5}$ (or $M_{\rm peak} < 10^{9}~\rm{M_{\odot}}$), which is below the suggested resolution limit of $N_{\rm peak} \sim 10^{3.5}$ as in~\cite{2024ApJ...970..178M}, Symfind still performs well and finds $10\%$ more subhalos than RCT. For subhalos above the $N_{\rm peak} \sim 10^{3.5}$ limit, Symfind typically recovers $\sim 50\%$ more subhalos than RCT, especially for the most massive subhalos which $N_{\rm peak} > 10^{4.5}$. The extra subhalos that Symfind tracks mostly experience a significant mass loss due to tidal stripping after an early infall.

For SIDM subhalos, the differences between RCT and Symfind become smaller, and Symfind recovers fewer subhalos in every $N_{\rm peak}$ bin, as shown in the bottom panels of figure~\ref{fig:subov-group}. For overall populations, the disagreements are primarily subhalos that experience significant mass loss and have an early infall time; interestingly, these are the systems for which Symfind outperforms RCT in the Group-CDM simulation. Below the limit of $N_{\rm peak} \sim 10^{3.5}$, the disagreements account for $\sim 10\%$ of the SIDM subhalo population; in this regime, Symfind still recovers some extra subhalos. On the other hand, Symfind loses track of almost $30\%$ of the RCT catalog for the most massive subhalos in this simulation. We highlight Subhalo~10 examined in figure~\ref{fig:group-case} with a red circle. This subhalo is not losing the most mass and does not have the earliest infall time in this bin, suggesting that the tracking problems we demonstrated in section~\ref{subsec:coreform} do not only affect the most extreme cases.

\section{Conclusions}
\label{sec:con}

In recent years, several groups have developed modern particle-tracking subhalo finders that demonstrate high accuracy in CDM simulations. In this paper, we test whether one such algorithm, Symfind, achieves comparable accuracy in both CDM and SIDM simulations on MW and Group mass scales, relative to a more traditional phase-space-based finder, RCT (Rockstar+Consistent-Trees). We found that Symfind systematically outperforms RCT in our CDM simulations, but the results are mixed for SIDM models with large cross section amplitudes and strong velocity dependence. In particular, we identify novel challenges in tracking SIDM subhalos: DM self-interactions can effectively kick the tightly bound particles, on which Symfind relies, out of the subhalo center, making them more likely to be stripped
away in the tidal field.

To demonstrate the impact of the challenges on the performance of Symfind and RCT, we selected representative examples from the Group-CDM and -SIDM simulations (section~\ref{sec:subhalocore}). Overall, the interplay between DM self-interactions and tidal effects influences subhalo tracking in a more subtle and complex manner compared to the CDM case, and there are no simple criteria to determine which finder performs better in general. Nevertheless, we identified several general trends. For massive core-expansion subhalos and core-collapse subhalos that experience close or multiple pericentric passages, a significant fraction of core particles can be lost, and Symfind does not offer a clear advantage. On the other hand, for core-collapse subhalos with large pericentric distances or fewer, more distant pericentric passages, Symfind typically outperforms RCT.

We further compared subhalo mass functions constructed using the Symfind and RCT catalogs (section~\ref{sec:shmf}). Their relative performance in identifying subhalo populations depends on the SIDM model and the mass scale of the host halo. For the Group-SIDM70 simulation, where $\sigma_0/m=70~{\rm cm^2\,g^{-1}}$ and $w=120~{\rm km\, s^{-1}}$, Symfind and RCT have similar performance overall, and the former even outperforms mildly for SIDM subhalos in the inner region ($r<R_{\rm vir}/4$). In contrast, for the Group-SIDM simulation with $\sigma_0/m=147.1~{\rm cm^2\,g^{-1}}$ and $w=120~{\rm km\, s^{-1}}$, Symfind does not outperform RCT in all radial bins, particularly, for massive subhalos. As the velocity turnover $w$ increases, these subhalos develop larger density cores that are more prone to tidal stripping, making them increasingly difficult to track with Symfind due to the enhanced loss of core particles. However, for the MW simulation with the same GroupSIDM model, Symfind and RCT perform comparably overall, even for relatively massive subhalos. This is likely because most of these subhalos are in the collapse phase. For the MW-MilkyWay simulation with $\sigma_0/m=147.1~{\rm cm^2\,g^{-1}}$ and $w=24.33~{\rm km\, s^{-1}}$, Symfind outperforms RCT. We further investigated the influence of infall time and mass loss on tracking (section~\ref{sec:subppl}).  

Since the performance of subhalo finders depends on the underlying SIDM model, and there is no clear preference for solely using Symfind or RCT, we recommend using a combination of both methods to construct the most robust SIDM subhalo catalogs. In our subhalo mass function tests, we find that the combined catalogs contain $2\textup{--}15\%$ more subhalos than either method alone when considering the overall population, and $10\textup{--}30\%$ more subhalos in the inner region ($r < R_{\rm vir}/4$) of the host system.

Our work can be further extended in several directions. For example, the evaporation effect~\cite{2025PhRvD.111f3001Z,Hainje:2025wtx}, which is caused by DM interactions between subhalo and host particles, can also contribute to tightly-bound particles being kicked out of subhalo centers~\cite{2016MNRAS.461..710D}. Isolating this effect is a fruitful area for future work. Additionally, it would be interesting to apply particle-tracking-based finders to SIDM simulations with a more exotic dark sector, such as including dissipative interactions~\cite{Huo:2019yhk,Shen:2021frv,Shen:2022opd,ONeil:2022szc,Roy:2024bcu,Leonard:2024mqo,Gemmell:2023trd,Shen:2025evo} or mass segregation effects~\cite{Patil:2025nmj,Yang:2025dgl,Yang:2025xsp}, which may affect the evolution of core particles.  

Furthermore, recent studies suggest that SIDM can introduce numerical problems in N-body simulations beyond those encountered in CDM~\cite{Zhong:2023yzk,2024A&A...689A.300F, 2024JCAP...09..074P,2024arXiv240201604M,Fischer:2025rky}. In particular, differences in gravothermal evolution predictions can arise from the different SIDM scattering implementations, time step sizes, particle numbers, as well as energy non-conservation in the deep core-collapse regime, which may further complicate the task of halo-finding algorithms. Although the loss of core particles may not be sensitive to the specific scattering algorithm, a side-by-side comparison study across different simulation software may provide insights into the most robust SIDM subhalo identification and tracking algorithms.

The loss of core particles has a more significant implication on using “orphan” models in SIDM simulations, which rely on one or more most-bound particles represent subhalos~\cite{1995ApJ...454....1S, 2006MNRAS.371..537W, 2010MNRAS.404.1111G, 2010ApJ...710..903M, 2019ApJS..245...16H, 2021ApJ...913..109S, 2021ApJS..252...19H,Bhattacharyya:2021vyd, 2023OJAp....6E..24K}, as such assumptions may not hold in SIDM. We note that it will be essential to develop similar tests for hydrodynamic simulations in both CDM and SIDM. {For example, a recent study shows that collisional particles (such as gas particles in hydrodynamic simulations) may not be reliable tracers for tracking subhalos~\cite{Moreno:2025nps}.} Furthermore, non-gravitational physics (e.g., supernova feedback) can presumably also lead to the unbinding of core particles in these scenarios, which we will leave for future work.


\acknowledgments

We thank Benedikt Diemer and Hyunsu Kong for useful discussion. This work used data from the \textsc{SIDM Concerto} simulations~\cite{Nadler:2025jwh}, which are publicly available on Zenodo at \href{https://doi.org/10.5281/zenodo.14933624}{DOI: 10.5281/zenodo.14933624}. HBY was supported by the U.S. Department of Energy (de-sc0008541) and the John Templeton Foundation (63599). The computations presented here were conducted through Carnegie's partnership in the Resnick High Performance Computing Center, a facility supported by Resnick Sustainability Institute at the California Institute of Technology. Computations were also performed using the computer clusters and data storage resources of the HPCC at UCR, which were funded by grants from NSF (MRI-2215705, MRI-1429826) and NIH (1S10OD016290-01A1). The opinions expressed in this publication are those of the authors and do not necessarily reflect the views of the funding agencies.

\bibliographystyle{JHEP}
\bibliography{biblio.bib}

\appendix

\section{Density Profile of Representative Examples}
\label{appex:sub-rho}

\begin{figure*}[p]
    \centering
    \includegraphics[width=0.83\columnwidth]{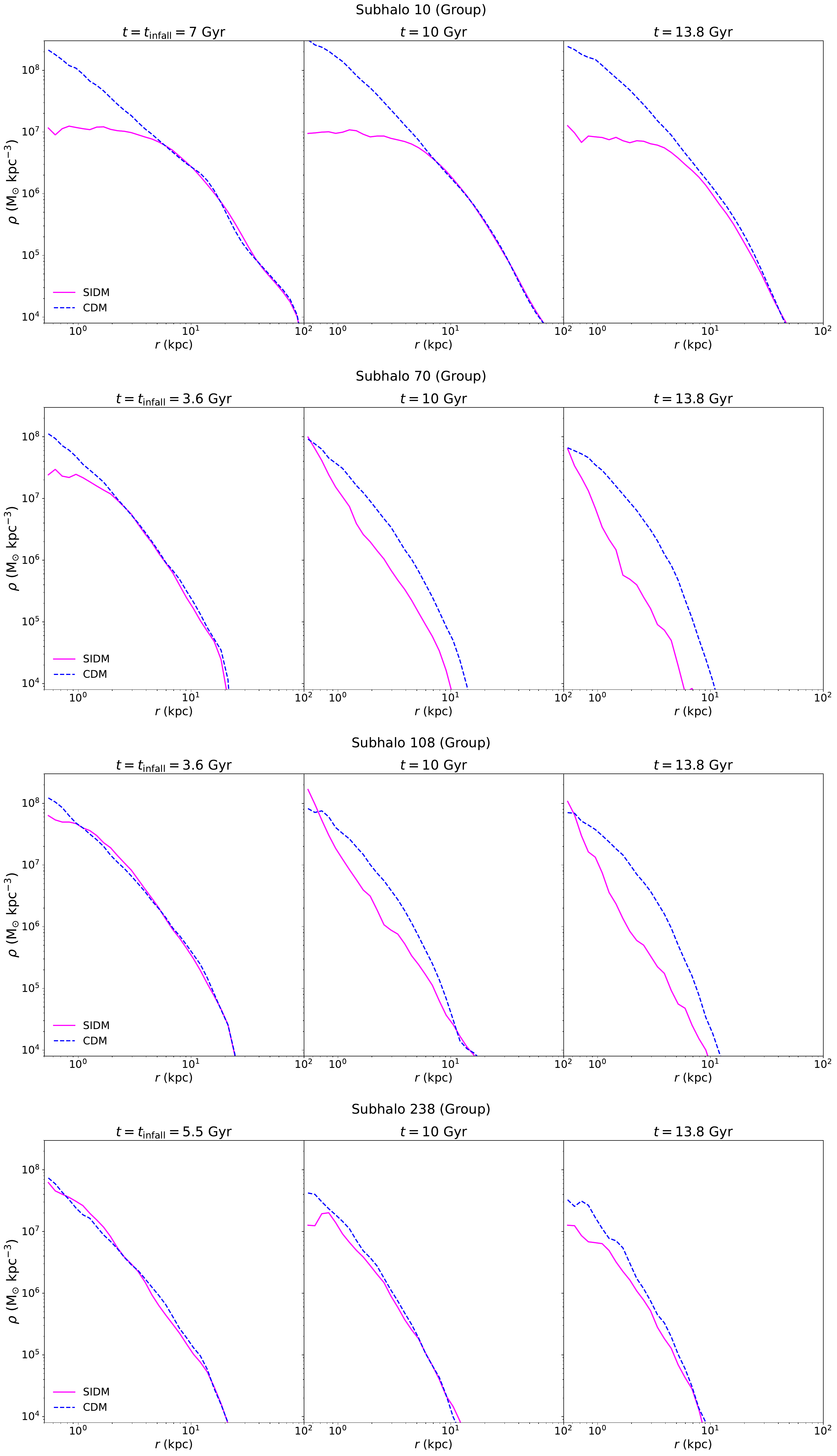} 
    \caption{Density profiles of representative examples from the Group CDM (dashed blue) and SIDM (solid magenta) simulations. The tidal evolution histories of these examples are discussed in detail in section~\ref{sec:subhalocore}. }
    \label{fig:all-rho}
\end{figure*}

In figure~\ref{fig:all-rho}, we present the density profiles of the representative examples discussed in section~\ref{sec:subhalocore}. The first row shows the CDM (dashed blue) and SIDM (solid magenta) density profiles for Subhalo~10 at three different snapshots $t=t_{\rm infall}=7~\rm{Gyr}$, $10~{\rm Gyr}$, and $13.8~{\rm Gyr}$. The SIDM halo is in the core-expansion phase at its infall and remains in that phase over the entire evolution period. The SIDM halo has a higher mass loss rate compared to its CDM counterpart, but their density profiles in the outer regions only differ mildly. The detailed tidal evolution of Subhalo~10 is shown in figure~\ref{fig:group-case}.

The second row shows Subhalo~70, which has an early infall time while a large pericenter distance as indicated in figure~\ref{fig:group-cc-lost-case-70}. At the infall, the SIDM halo still has a $1~{\rm kpc}$ density core, but it evolves deeply into the collapse phase at the later stages, after experiencing a significant mass loss. At $t=13.8~{\rm Gyr}$, the SIDM density profile scales as $\rho\propto r^{-4}$. The third row shows Subhalo~108, which is similar to Subhalo~70, but with a smaller pericenter as indicated in figure~\ref{fig:group-cc-lost-case-108}. The fourth row shows Subhalo~238. The SIDM halo is deep in the collapse phase at the infall, and it suffers a mild mass loss compared to the CDM counterpart. The detailed tidal evolution of Subhalo~238 is shown in figure~\ref{fig:group-cc-lost-case-238}.

\section{Evolution of Core Particles in Idealized Simulations}
\label{appex:core-ideal}

\begin{figure*}[h]
    \centering
    \includegraphics[width=\columnwidth]{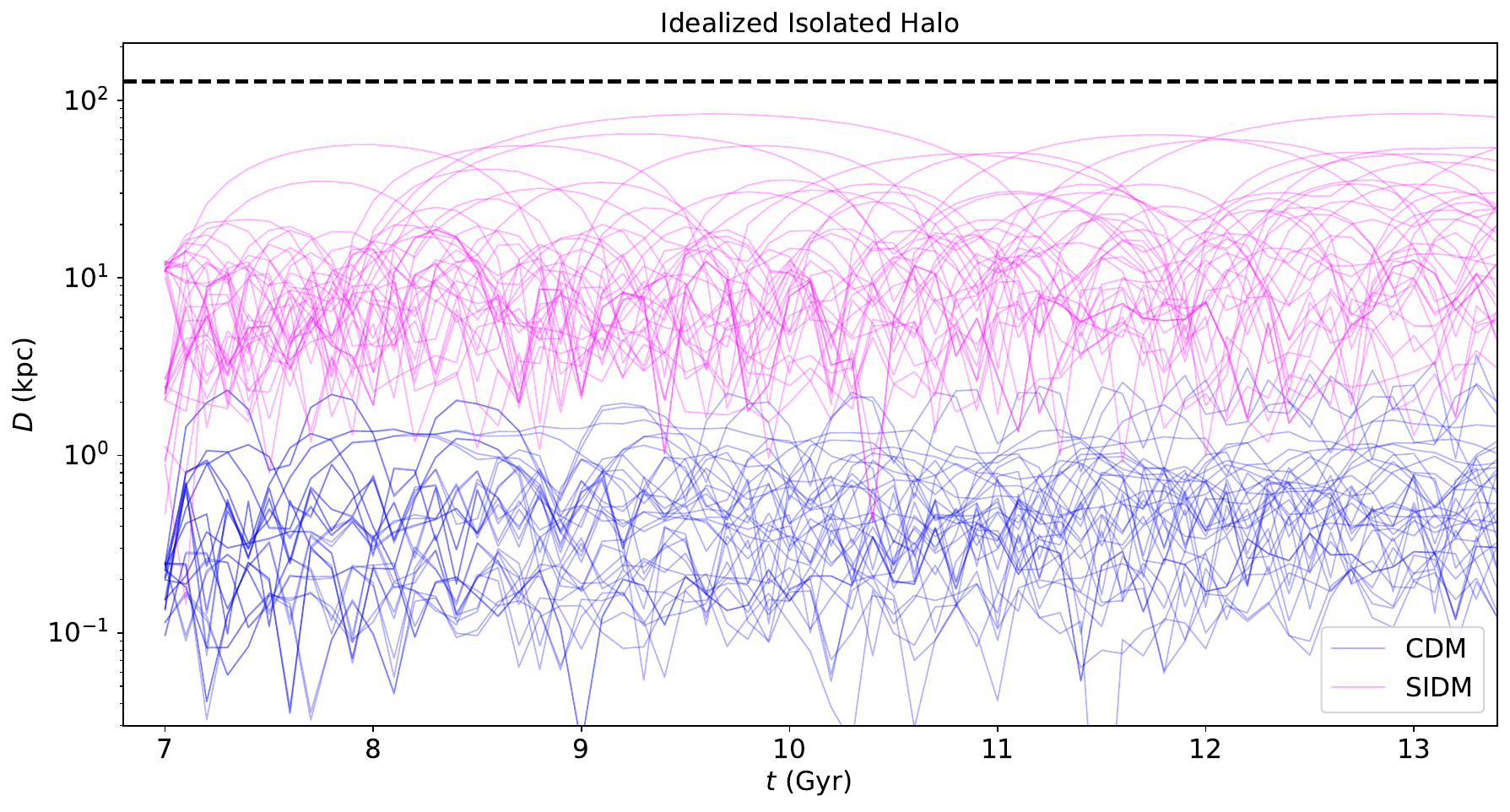} 
    \caption{Evolution of the distance of core particles from the halo center in idealized CDM (blue) and SIDM (magenta) simulations. Core particles are tagged at $7~{\rm Gyr}$. The dashed black line indicates the halo's virial radius.}
    \label{fig:iso-pos-ideal}
\end{figure*}

We conduct idealized N-body simulations for isolated CDM and SIDM halos to test our findings about the ``diffusion" behavior of the core particles caused by DM self-scattering. We use the public code SpherIC~\cite{2013MNRAS.433.3539G} to generate the initial condition and the code GADGET-2~\cite{2001NewA....6...79S, 2005MNRAS.364.1105S}, implemented with an SIDM module~\cite{2022JCAP...09..077Y,Yang:2020iya} to perform simulations. The same SIDM module was used for performing our zoom-in cosmological simulations. We assume that the initial halo follows the standard Navarro–Frenk–White density profile~\cite{1997ApJ...490..493N}, with a scale density of $\rho_{s} = 6.5\times 10^{6}~\rm{M_{\odot}/kpc^3}$ and a scale radius of $r_{s} = 9.3 ~\rm{kpc}$, corresponding to a halo mass of $10^{11}~\rm{M_{\odot}}$ and a median concentration. We set the values of the particle mass, softening length, and SIDM model to match those used in the Group simulations, as described in section~\ref{subsec:simov}.

In figure~\ref{fig:iso-pos-ideal}, we show the distance of core particles from the halo center in idealized CDM (blue) and SIDM (magenta) simulations. In the CDM halo, nearly all core particles remain within $1~{\rm kpc}$ of the center at all times. In contrast, core particles in the SIDM halo are dispersed over a much wider range. Most of them can migrate to $20~{\rm kpc}$ and a few of them even reach $80\textup{--}100~{\rm kpc}$. Since tidal stripping is absent in the isolated simulations, the difference in the distribution of core particles between SIDM and CDM halos is entirely due to DM self-interactions. We have also verified that the core size of the simulated SIDM halo is $20~{\rm kpc}$, consistent with the migration range of the core particles.

\end{document}